# 3D orientation super-resolution spatial-frequency-shift microscopy

Xiaowei Liu,[a, †] Mingwei Tang,[b, †] Ning Zhou,[c] Chenlei Pang,[a] Zhong Wen,[b] Xu Liu,[b, d, e] and Qing Yang[a, b, d, e, *]

[a] Research Center for Humanoid Sensing, Zhejiang Lab, Hangzhou 311121, China
[b] State Key Laboratory of Extreme Photonics and Instrumentation, College of Optical Science and Engineering, Zhejiang University, Hangzhou 310027, China
[c] School of Physics and Optoelectronic Engineering, Hangzhou Institute for Advanced Study, University of Chinese Academy of Sciences, Hangzhou 310024, China
[d] ZJU-Hangzhou Global Scientific and Technological Innovation Center, Hangzhou 311215, China
[e] Collaborative Innovation Center of Extreme Optics, Shanxi University, Taiyuan, China
† These authors contributed equally to this work.

**Abstract:** Super-resolution mapping of the 3D orientation of fluorophores reveals the alignment of biological structures where the fluorophores are tightly attached, and thus plays a vital role in studying the organization and dynamics of bio-complexes. However, current super-resolution imaging techniques are either limited to 2D orientation mapping or suffer from slow speed and the requirement of special labels in 3D orientation mapping. Here, we propose a novel polarized virtual spatial-frequency-shift effect to overcome these restrictions to achieve a universal 3D orientation super-resolution mapping capability. To demonstrate the mechanism, we simulate the imaging process and reconstruct the spatial-angular information for sparsely distributed dipoles with random 3D orientations and microfilament-like structures decorated with fluorophores oriented parallel to them. The 3D orientation distribution can be recovered with a doubled spatial resolution and an average angular precision of up to 2.39 degrees. The performance of the approach with noise has also been analyzed considering real implementation.

## 1. Introduction

The extraction of 3D orientation angular information of the alignment and rotational movements of molecules plays a vital role in studying the organization and dynamics of bio-complexes, and thus can push forward the progress of life science, cytology, etc. (Fig. 1a). For example, by directly measuring the tilt and wobble of individual fluorescent probes, researchers have gained an understanding of the conformations of DNA[1-3], the motions of molecular motors[4], the organization of molecular assemblies in cell membranes [5, 6], as well as the structure of biological filaments[2, 7, 8]. Consequently, the study on orientation mapping microscopy has attracted interest worldwide.

In recent years, myriads of 2D orientation microscopy methods (2D projection of the orientation on the imaging plane) have been reported [7, 9, 10]. Generally speaking, in the emission-polarization-splitting scheme[11], the 2D angular information can be inferred using ratio-metric intensity measurements between different polarization channels. In the illumination-polarization-modulation scheme, the 2D angular information is calculated based on the Cosine quantitative dependence of the absorption efficiency on the intersection angle between the excitation electric field and the dipole orientation. By introducing polarized structured illumination, the 2D orientation mapping with super-resolution has been achieved [7, 12] (Fig. 1b).

Compared with 2D orientation microscopy, 3D orientation microscopy provides more information for understanding the biological and chemical processes. For instance, cell membranes are characterized by the 3D organization and alignment of their lipid constituents; the 3D folding conformation or structural disorder of a protein largely determines its interactions with neighbors. One method to sense the 3D orientation is inferring the angle through a series of defocus images[13]. Nevertheless, the spreading defocus point spread function (PSF) makes the spatial resolution far below the Abbe diffraction limit, and the 3D orientation angular information would be mixed between two diffraction-limited dipoles (Fig. 1c).

To increase the resolution in 3D orientation mapping, wide-field single-molecule localization microscopy (SMLM) beyond the diffraction limit is utilized [14-17]. By encoding the 3D orientation information into the shape of PSF, the location and orientation of stochastically activated fluorescent molecules can be simultaneously reconstructed from diffraction-limited image sequences, with ultra-high resolution up to sub-20nm. The main concern regarding these SMLM-based orientation mapping methods may be the dependence on the special labels with blinking properties, and the difficulty to achieve fast dynamic imaging for thousands of raw frames that need to be captured. Consequently, visualizing the 3D orientation with super-resolution in a high-throughput and universal scheme remains to be explored.

The spatial-frequency-shift effect stands out for its compatibility with both labeled and label-free imaging[18-23], as well as high-throughput data extraction capability in the reciprocal Fourier space [24, 25], which may overcome the restrictions faced by current 3D orientation microscopy. Here, we propose a novel polarized virtual spatial-frequency-shift microscopy (**PVSFSM**) for 3D orientation super-resolution imaging compatible with normal fluorescence labeling, with resolution improved twice the Abbe diffraction limit. An equivalent cosine structured illumination pattern is generated to introduce the spatial-frequency-shift effect for super-resolution imaging by scanning an intensity-modulated tight focus with varied x, y, z polarized components [26, 27] in the sample domain (Fig. 1d). The emission carries mixed high spatial frequency spectrums outside the passband of objective regarding different orientation projection. The mixed spatial-angular information can be decoupled in the reciprocal domain by processing the raw images captured with different illumination polarization state at the back aperture of the objective. Finally the 3D orientation distribution can be reconstructed beyond the diffraction limit (Fig. 1e).

We demonstrate this 3D-orientation PVSFSM approach in simulation for sparsely distributed dipoles with random 3D orientations and microfilament-like structures decorated with fluorophores oriented parallel with it. Resolutions of $0.19\lambda$ and $0.44\lambda$ can be achieved in lateral and axial dimensions respectively, and the 3D orientation recovering precision reaches 2.39 degrees for the random distributed dipole sample and 2.01 degrees for microfilament-like sample. The influence of Gaussian noise on the spatial resolution and orientation precision has also been analyzed. This research will benefit nanoscale structural imaging in fundamental biology, medicine, pathologies, material science, etc.

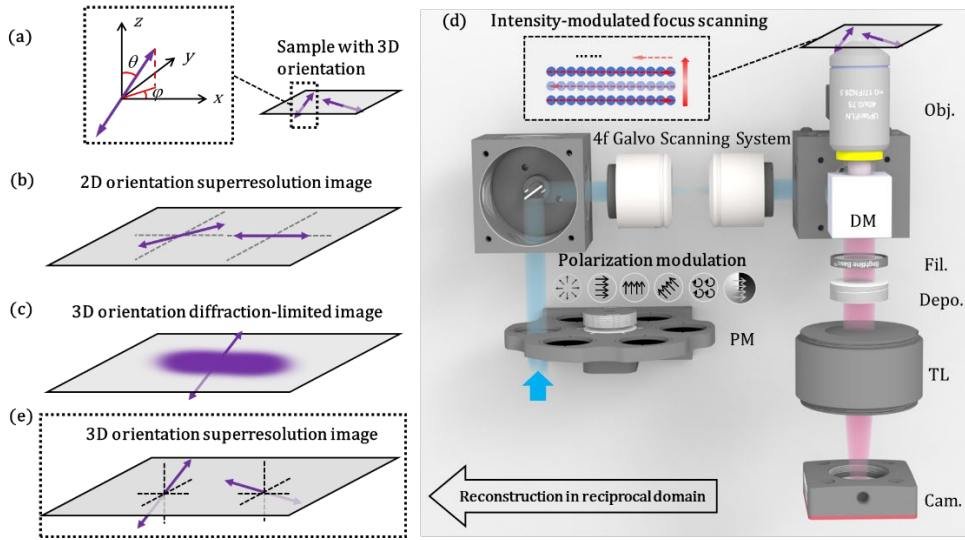

**Fig. 1. (a)** Schematic of a sample with 3D orientation. The coordinate shows the 3D orientation of a dipole can be described by the polar angle θ and azimuthal angle φ. **(b)** 2D orientation super-resolution image. Only the projection of the orientation on the imaging plane can be mapped. **(c)** 3D orientation diffraction-limited image. The orientation information of two diffraction-limited dipoles are mixed. **(d)** Schematic setup of the 3D-orientation PVSFSM. The focused light spot is intensity modulated when scanning in the focus plane to generate an equivalent structured illumination, and the camera keeps on during the scanning and captures one image at the end. PM: polarization modulator, Obj: objective lens, DM: dichroic mirror, Fil: bandpass filter, Depo: depolarizer, TL: tube lens, Cam: camera. **(e)** 3D orientation super-resolution image, in which the complete 3D orientation information can be reconstructed.

## 2. Principle

The emission rate of a dipole is proportional to the intensity of the projection of the in-situ electric field $E$ on the orientation of the dipole. The emission rate distribution of a sample consisting of variously oriented molecules under a polarized illumination can be expressed by Eq. 1.

$$\begin{aligned}
\text{Emis.} &= S\left|E_x \sin\theta\cos\varphi + E_y \sin\theta\sin\varphi + E_z \cos\theta\right|^2 + D\left|E_x + E_y + E_z\right|^2 \\
&= E_x E_x^*(S\cos^2\varphi\sin^2\theta + D) + E_y E_y^*(S\sin^2\varphi\sin^2\theta + D) \\
&\quad + E_z E_z^*(S\cos^2\theta + D) + (E_x E_y^* + E_x^* E_y)S\cos\varphi\sin^2\theta\sin\varphi \\
&\quad + (E_x E_z^* + E_x^* E_z)S\cos\varphi\sin\theta\cos\theta + (E_y E_z^* + E_y^* E_z)S\sin\varphi\sin\theta\cos\theta \\
&= \sum_{m}^{1,2,\ldots 6} Illu_m S_m
\end{aligned}$$ 
Eq. 1

$Illu_m$ and $S_m$ are defined as follows.

$$\begin{aligned}
Illu_1 &= E_x E_x^* & S_1 &= S\cos^2\varphi\sin^2\theta + D \\
Illu_2 &= E_y E_y^* & S_2 &= S\sin^2\varphi\sin^2\theta + D \\
Illu_3 &= E_z E_z^* & S_3 &= S\cos^2\theta + D \\
Illu_4 &= E_x E_y^* + E_x^* E_y & S_4 &= S\cos\varphi\sin^2\theta\sin\varphi \\
Illu_5 &= E_x E_z^* + E_x^* E_z & S_5 &= S\cos\varphi\sin\theta\cos\theta \\
Illu_6 &= E_y E_z^* + E_y^* E_z & S_6 &= S\sin\varphi\sin\theta\cos\theta
\end{aligned}$$

In Eq.1, $\varphi$ is the azimuthal angle, and $\theta$ is the polar angle of the molecule orientation, as diagrammed in the coordinate in Fig. 1a. S describes the emission rate of the sample with polarization property; and D describes the emission rate of the sample without polarization property. Consequently, the spatial distribution of a sample consisting of variously oriented molecules can be decomposed into 6 components: $S_m$ (m = 1, 2, …6), which are different functions of the azimuthal and polar angle distribution of the sample, and respond to different components of the illumination (i.e., $Illu_{1, 2, \ldots 6}$) in the contribution to the emission rate. Given the distribution of the 6 components of the illumination well known, it is possible to reconstruct $S_m$ (m = 1, 2, …6) and calculate the distribution of $\theta$ and $\varphi$.

To enable super-resolution observation of the 3D orientation without utilizing special labeling, structured polarized illumination is required to shift the high spatial frequency spatial-angular information to the detection band. In the proposed 3D-orientation PVSFSM, the structured illumination is generated equivalently by rapidly scanning a polarization-modulated tight focus in the sample domain, as shown by the schematic in Fig. 1d. NA of the objective is large enough to meet the tight focus condition, so as to stimulate an axially polarized electric field in the focus, which is essential for 3D orientation mapping. Consequently, a tight focus consists of 6 components $Illu_{1, 2, ...6}$ defined in Eq. 1. A depolarizer is placed in the detection path to make an orientation-irrelevant PSF. In such a scheme, the final captured image can be expressed as Eq. 2.

$$\begin{aligned} Image(r') &= \sum_{m}^{1,2,...6} \iint M(\bar{r})PSF_m^{illu.}(r-\bar{r})S_m(r)PSF^{det.}(r-r')drd\bar{r} \\ &= \sum_{m}^{1,2,...6} \int [\int M(\bar{r})PSF_m^{illu.}(r-\bar{r})d\bar{r}]S_m(r)PSF^{det.}(r-r')dr \\ &= \sum_{m}^{1,2,...6} \{[M(r') \otimes PSF_m^{illu.}(r')]S_m(r')\} \otimes PSF^{det.}(r') \end{aligned}$$

Eq. 2

We use the subscripts "*illu.*" and "*det.*" in the following text to represent the quantity involving illumination and detection, respectively. In Eq. 2, we assume the magnification factor of the microscope system is 1 for clarity. $r$ and $r'$ are the coordinates in the domain of object and image, respectively. The term $Illu_m$ in Eq. 1 has been substituted by $PSF_m^{illu.}$ in Eq. 2, which denotes the $m$'th components of the tight focus in illumination. $PSF^{det.}$ denotes the PSF of detection (which is irrelevant to the emission polarization due to the depolarizer placed in the detection path). $\otimes$ denotes convolution. $\bar{r}$ denotes the center position of the scanning tight focus, whose intensity is modulated over $\bar{r}$: $M(\bar{r}) = 1 + \cos(2\pi \boldsymbol{f_0} \cdot \bar{r} + \psi)$, in which $\boldsymbol{f_0}$ is the spatial frequency of the cosine structured pattern; The phase $\Psi$ determines the position of the pattern. The term $M(\boldsymbol{r'}) \otimes PSF_m^{illu.}(\boldsymbol{r'})$ plays the role of the structured illumination field as in wide-field SIM [28]. The equivalent structured illumination pattern changes for different $S_m$.

In the reciprocal space, Eq.2 can be expressed by Eq. 3.

$$\begin{aligned} Spectrum(f') &= \sum_{n}^{-1,0,1} \left( \sum_{m}^{1,2...6} 0.5^{|n|} \exp(in\Psi)\delta(f'-nf_0)OTF_m^{illu.} \otimes F_m OTF^{det.} \right) \\ &= \sum_{n}^{-1,0,1} \left( \sum_{m}^{1,2...6} 0.5^{|n|} \exp(in\Psi)OTF_m^{illu.}(nf_0)F_m(f'-nf_0)OTF^{det.} \right) \end{aligned}$$

Eq. 3

In Eq.3, $OTF_m^{illu.}$ is the Fourier transform of $PSF_m^{illu.}$, i.e. the spatial frequency spectrum of the m'th component of the tight focus. $OTF^{det.}$ is the Fourier transform of $PSF^{det.}$, i.e. the optical transfer function of the detection. $F_m$ denotes the spatial frequency spectrum of $S_m$. Similar to the conventional SIM, the spatial frequency spectrum can be decomposed into 3 components with different frequency shift vectors (FSVs) corresponding to $nf_0$, n = -1, 0, 1. Differently, each component with one FSV contains 6 sub-components coming from $S_m$ (m=1, 2, …6), whose weight is determined by the $OTF_m^{illu.}$ of the present tight focus at the corresponding FSV, as illustrated in Fig. 2a. All the shifted sub-components are filtered by the low-pass optical transfer function of the objective $OTF^{det.}$ and summed to form the final image (Fig. 2a).

Consequently, the spatial-angular components $S_m$ (m=1, 2, …6) mixed in the real space can be well separated in the reciprocal space. To solve the 18 sub-components $F_m(f' - nf_0)OTF^{det.}$ (m=1, 2…6; n=-1, 0, 1) in Eq. 3, different Ψ (to change exp(inΨ) in Eq. 3) and different tight focus (to change $OTF_m^{illu.}$ in Eq. 3) should be utilized to provide multiple independent combinations of these unknown sub-components in the reciprocal domain. We use 3 phases Ψ (Ψ=0, 2/3π, 4/3π) and select 6 different polarization states in the illumination path (as illustrated in Fig. 1d) at the aperture plane of the objective to form various $OTF_m^{illu.}$, to establish an equation set with the full rank coefficient matrix. The 6 polarization states include 3 linearly polarization states along 0, 45, and 90 degrees with the x-axis, left circularly polarization state, radial polarization state, and linear polarization state with a vortex phase of topological charge of 1. For most FSVs, these 6 polarization states can provide orthogonal coefficients $OTF_m^{illu.}(nf_0)$. As shown in Fig.2b, for various polarization state in the aperture plane, the formed tight focus possess quite different 6 components in the real domain ($PSF_{1,2...6}^{illu.}$) and in the reciprocal domain ($OTF_{1,2...6}^{illu.}$). The simulation on the tight focus is done assuming the objective lens satisfies sine condition [26].

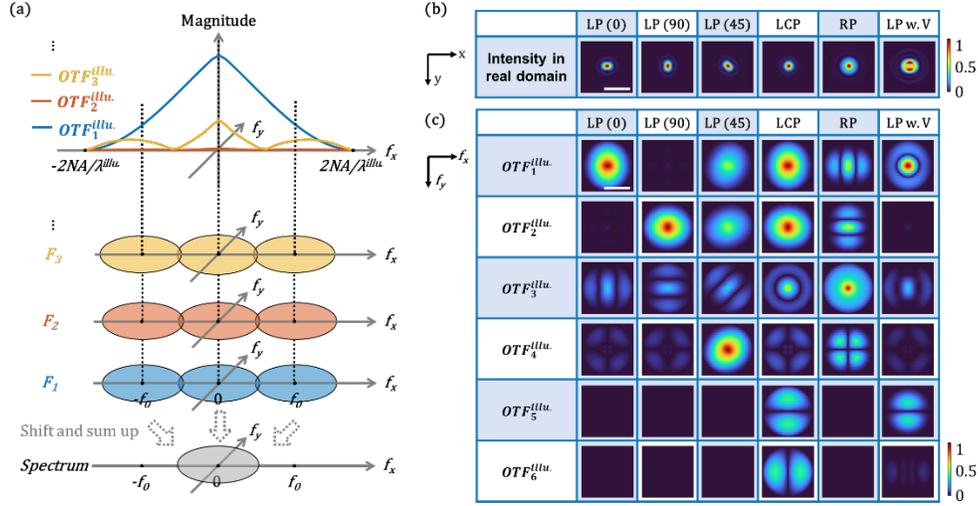

**Fig. 2. (a)** Decomposition of the captured PVSFSM image in the reciprocal space. $F_m$ (m = 1, 2, 3…) represent the spatial frequency spectrum of $S_m$, consisting of the 3D orientation information of the sample, as defined in Eq.1. $OTF_m^{illu.}$ (m = 1, 2, 3…) represent the spatial frequency spectrum of $PSF_m^{illu.}$, different components of the tight focus. Each sub-component of $F_m$ is shifted to the center in the reciprocal domain and summed up with a weight equals to $OTF_m^{illu.}$ at the center position of the sub-component. **(b-c)** Spatial distribution of the total intensity (b) and spatial frequency spectrum of different components (c) of the tight focus corresponding to different polarization states at the back aperture of the objective, denoted at the top row. LP (0): linear polarization with 0 degrees. LP (90): linear polarization with 90 degrees. LP (45): linear polarization with 45 degrees. LCP: left circular polarization. RP: radial polarization. LP w. T: x linear polarization state with a vortex phase of topological charge of 1. Scale bar in (b): 1 μm. Scale bar in (c): $NA/\lambda^{illu.}$. NA = 1.4. $\lambda^{illu.}$ = 561nm.

The resolved $F_m(f' - nf_0)OTF^{det.}$ can be shifted back in the reciprocal space and stitched together for the same *m* to enable a broader detection range for super-resolution microscopy. Six super-resolution images of $S_m$ (m=1, 2, …6) can be achieved, and then the azimuthal angle $\varphi$ and polar angles $\theta$ can be calculated with the same spatial resolution from $S_m$ (m=1, 2, …6) using an optimization Gaussian-Newton algorithm. The maximum frequency shift value is limited by the passing band of the illumination objective, *i.e.*, $2NA/\lambda^{illu.}$, as illustrated in Fig. 2a. Consequently, the resolution limit in 3D-orientation PVSFSM can be calculated by $\lambda^{det.}/(2NA(1+\lambda^{det.}/\lambda^{illu.}))$, slightly more than 2 times surpassing the Abbe diffraction limit.

## 3. Results

*3.1 3D orientation mapping with lateral super-resolution*

Firstly, we generate a 2D sample of sparsely distributed dipoles with random 3D orientations to illustrate the 3D orientation reconstruction process with enhanced lateral resolution (noise-free). We generate 27 dipoles with varying 3D orientations as shown in Fig. 3a. All the dipoles possess a 10% baseline which bears no polarization dependence. Then the sample is scanned with a 561-nm focused illumination light using an oil-immersed 1.4-NA objective to excite fluorescence. The fluorescence emission is then collected with the same objective; passes through a 640 nm long-pass spectral filter, a depolarizer, a tube lens, and finally is captured by a CCD camera. Six different polarization states are used to generate the tight focus for illumination, as introduced in Fig. 1d and Fig.2b. For each polarized state, the tight focus is modulated in intensity according to its location to form a virtual structured illumination pattern with pre-defined spatial frequency and phase. The frequency shift distances are selected to be $NA/\lambda^{det.}$ and $2NA/\lambda^{det.}$. The frequency shift direction is along the 45, and 135 degrees with x-axis for $NA/\lambda^{det.}$, and along 22.5, 67.5, 112.5, and 157.5 degrees with x-axis for $2NA/\lambda^{det.}$. Consequently, 6 varied cases are included here regarding the periods and directions of the structured illumination. For one FSV, the polarization of the illumination is modulated with 6 states at the back aperture of the objective, and the phase $\Psi$ is modulated with 3 values, i.e. 18 raw images would be captured (as illustrated in Fig. 3b), to provide 18 equations with a well-conditioned coefficient matrix to solve the 18 sub-components $F_m(f' - n\boldsymbol{f_0})OTF^{det.}$ (m=1, 2…6; n=-1, 0, 1).

After solving the sub-components of the spatial spectrum of $S_m$ (m=1, 2, …6) from the linear equation sets corresponding to all the FSV and stitching them together, the spatial spectrums of $S_m$ (m=1, 2, …6) achieve two times wider detection range than the passband of the objective. The reconstructed super-resolution images of $S_m$ (m=1, 2, …6) are shown in Fig. 3c1-c6. Then we use the Gaussian-Newton method to reconstruct the distribution of the azimuthal and polar angles from the super-resolution images of $S_m$ (m=1, 2, …6). Fig. 3d shows the 3D vector plot for the reconstructed result, which is close to the ground truth shown in Fig. 3a. Figs. 3e-f show the reconstructed and the true azimuthal and polar angles for the 27 dipoles. To quantitatively assess the 3D orientation mapping accuracy, we perform a statistical analysis for 135 dipoles about the angular deviation between the reconstruction and the ground truth (as shown in Fig. 3g). The reconstruction error is 2.39±2.46 degrees. Fig. 3h plots the imaging intensity distribution across two closely placed dipoles at the center of the field of view (shown by the yellow dash line in the magnified inset in Fig. 3c1-c3) of $S_m$ (m = 1, 2, 3) and $\sum_{m}^{1,2,3} S_m$ (the physical meaning of which is the emission rate when the sample is excited by a non-polarized light field). The two dipoles with 120 nm (0.19λ) center-to-center distance are clearly distinguished. Moreover, the 3D orientations of the two diffraction-limited located dipoles are also precisely reconstructed, as shown in Fig. 3e-f (No. of the two diffraction-limited located dipoles are 1 and 2), demonstrating a 3D orientation mapping resolution up to 0.19λ.

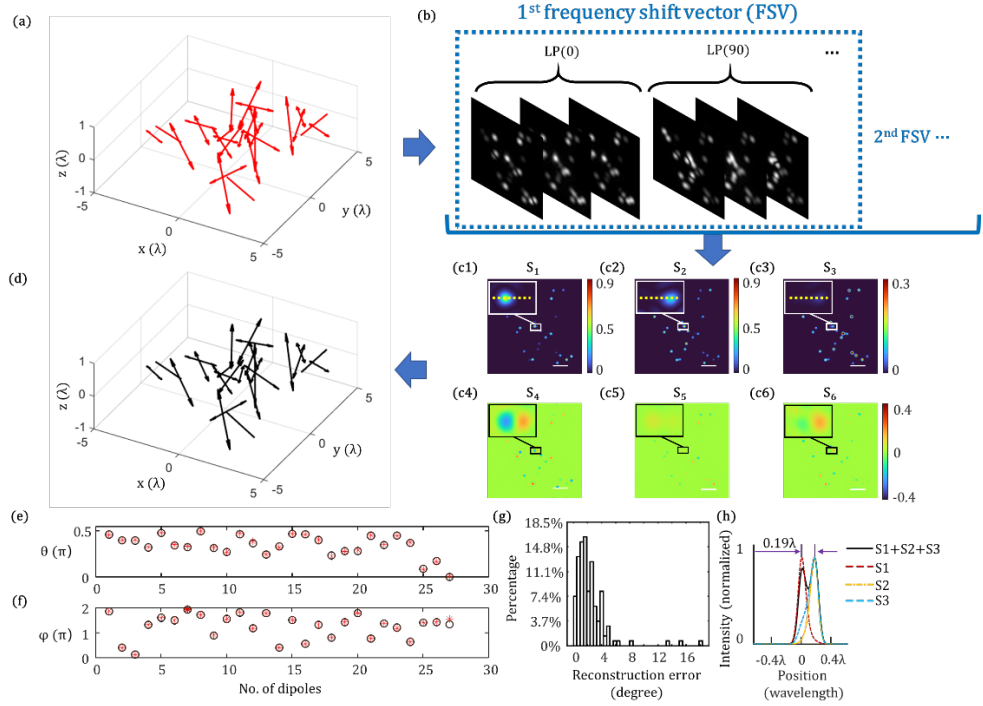

**Fig. 3**. 3D orientation super-resolution mapping by PVSFSM at noise-free condition. **(a)** 3D vector plot of the ground truth of the 3D orientation of sparsely distributed dipoles. **(b) Schematic of** raw images corresponding to different FSV and polarized illumination. **(c1-c6)** Reconstructed image of $S_m$ (m=1, 2, …6) with lateral super-resolution. The inset in the top left shows a magnified image of two selected 120 nm (0.19λ) spaced dipoles with different 3D orientations. **(d)** 3D vector plot of the reconstructed 3D orientation. **(e-f)** The reconstructed (red star) and ground truth (dark circle) of the polar (e) and azimuthal (f) angles of the 27 dipoles. **(g)** The statistical reconstruction error of 3D orientation angle. Statistical analysis was performed on 135 points. **(h)** The line scan of the imaging intensity along the yellow dash lines in (c1-c3), demonstrating a distance of 120 nm ($0.19\lambda^{dec.}$) can be resolved. λ in the figure denotes $\lambda^{det.}$. NA = 1.4. $\lambda^{dec.}$ = 640 nm.

*3.2 Results with noise*

The noise would affect the reconstruction process and degrade the resolution, contrast, and angle precision in practice. In 3D-orientation PVSFSM, the weight of the shifted spatial spectrum of the object is equal to the spatial spectrum of the illumination tight focus at the corresponding FSV (Fig. 2a). Consequently, when the shift distance is close to $2NA/\lambda^{illu.}$, the weight of the shifted spatial frequency spectrum is near to 0, and thus would be more easily submerged by the noise. Consequently, in the noisy case, we need to select a smaller frequency shift distance compared with the noise-free case, due to which the final resolution decreases.

To investigate the influence of noise on the 3D-orientation PVSFSM, we further conducted simulations with Gaussian noise added, which is a typical comprehensive noise. The SNR of the raw images is set to be 13dB. The frequency shift distances are selected to be $NA/\lambda^{det.}$, $1.7NA/\lambda^{det.}$ along the 45 and 135 degrees with x axis. The maximum of the reconstructed spatial frequency is $3.7NA/\lambda_{det.}$, slightly lower than the noise-free case. The raw images corresponding to the frequency shift distance of $2NA/\lambda^{det.}$ are also included in the simulation to provide redundant data for denoising, while only the spatial frequencies within $3.7NA/\lambda_{det.}$ that are overlapped by the detection ranges of at least two FSV are trusted to calculate the final super-resolution image, as demonstrated in Fig. 4a. We also use the sparsely distributed dipoles with random 3D orientation as the sample, as shown in Fig. 4b. Fig. 4c shows several raw diffraction-limited images captured with linearly polarized illumination (along the x-axis) to manifest the level of SNR. The spatial spectrum is reconstructed with a regress method to get use of the redundant data in the overlapped region provided by different FSVs. The azimuthal and polar angles are then calculated based on the super-resolution image of $S_m$ (m=1, 2, …6) by the Gaussian-Newton method.

Fig. 4d and 4e show the reconstructed super-resolution image of $S_m$ (m=1, 2, …6) and the 3D vector plot of the reconstructed orientation of the dipoles. Fig. 4f and 4g show the reconstructed and the true azimuthal and polar angles of the dipoles. Compared with the noise-free case, a larger deviation between the reconstructed angular result and the ground truth can be observed, especially for the polar angle θ. The 3D orientation reconstruction angular deviation is 7.08±6.43 degrees (Statistical analysis was performed on 45 points). The line scans across two 160 nm spacing dipoles

(shown by the yellow dash line in Fig. 4c) of the intensity distribution of the reconstructed image of $S_m$ (m=1, 2, 3) and $\sum_i^{1,2,3} S_i$, show that the two dipoles can be distinguished in intensity. Besides, the reconstructed 3D orientation angles of the two dipoles are very close to the ground truths, demonstrating a resolution of $\lambda/4$ can be achieved in 3D orientation PVSFSM at 13dB SNR.

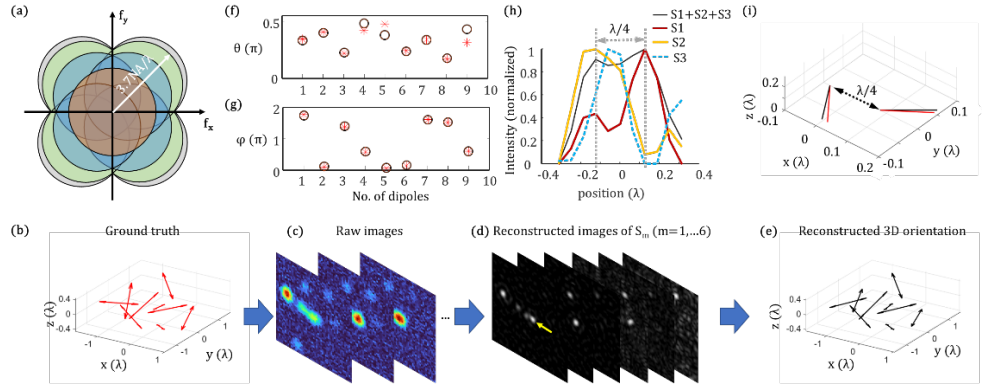

**Fig. 4** 3D orientation super-resolution mapping by PVSFSM at noisy conditions. **(a)** The detected range in the spatial frequency domain. Orange: no shift corresponding to the DC component of the structured illumination. Blue: frequency shift distance = $NA/\lambda^{det.}$; Green: frequency shift distance = $1.7NA/\lambda^{det.}$; Gray: frequency shift distance = $2NA/\lambda^{det.}$; Considering the usage of abundant data in the overlap region to suppress noise, the reconstructed maximum spatial frequency is $3.7NA/\lambda^{det.}$. **(b)** 3D vector plot of the ground truth of the 3D orientation of sparsely distributed dipoles. **(c)** Schematic of the raw images. **(d)** Reconstructed images of $S_m$ (m=1, 2, …6) with lateral super-resolution. **(e)** 3D vector plot of the reconstructed 3D orientation. **(f-g)** The reconstructed (red star) and ground truth (dark circle) of the polar (f) and azimuthal (g) angles of the 9 dipoles. **(h)** The line scan of the imaging intensity of $S_m$ (m = 1, 2, 3) across two 160 nm ($\lambda^{det.}/4$) spaced dipoles shown by the yellow arrow in (d). **(i)** The 3D vector plot of the true and reconstructed orientation of the two 160 nm ($\lambda^{det.}/4$) spaced dipoles. $\lambda$ in the figure denotes $\lambda^{det.}$. NA = 1.4. $\lambda^{det.}$ = 640 nm.

## 4. Discussion

*4.1 3D orientation mapping with 3D spatial super-resolution*

This 3D orientation super-resolution mapping method can also be adapted into the pipeline of 3D spatial-frequency-shift microscopy, which we call 3D-orientation-3D-PVSFSM. In 3D-

orientation-3D-PVSFSM, the position-related intensity modulation of the tight focus in the 3D sample domain is

$$M(r) = \left| e^{i2\pi/\lambda^{det.}(sin\alpha \cdot r_{xy} + cos\alpha \cdot z)} + e^{i\left(\frac{2\pi z}{\lambda^{det.}} + \Psi\right)} + e^{i[2\pi/\lambda^{det.}(-sin\alpha \cdot r_{xy} + cos\alpha \cdot z) + 2\Psi]} \right|^2,$$

corresponding to the intensity distribution of a three-beam interference (one center beam propagating along the axis and two symmetrical side beams propagating with an intersection angle of $\alpha$ to the optical axis). The generated illumination pattern is held fixed with the detection objective, and the 3D imaging data is achieved by capturing a series of 2D images when axially moving the object. Similar to the conventional 3D structured illumination microscopy with a real three beam interference, the Fourier composition of the 3D imaging data in the 3D PVSFSM includes 5 components in the reciprocal domain, which corresponds to different lateral frequency shift distance of $-\frac{2sin\alpha}{\lambda^{det.}}$, $-\frac{sin\alpha}{\lambda^{det.}}$, $0$, $\frac{sin\alpha}{\lambda^{det.}}$ and $\frac{2sin\alpha}{\lambda^{det.}}$ respectively (among which the detection OTF corresponding to $\pm\frac{sin\alpha}{\lambda^{det.}}$ consisted of two copies of the conventional OTF, offset by a distance of $\frac{1-cos\alpha}{\lambda^{det.}}$ above and below the $f_z$=0 plane in the reciprocal domain), and the phase $\Psi$ in the expression of M($r$) needs to change 5 times during the z-scanning process to separate these 5 components. Differently, each component in the 3D PVSFSM also contains 6 sub-components come from $S_m$ (m=1, 2, …6), whose weight is determined by the value of 3D $OTF_m^{illu.}$ at the corresponding FSV. Consequently, 6 different polarization state at the back aperture of the illumination are switched besides the switch of phase $\Psi$, to provide a well-conditioned coefficient matrix to separate the spectrum of $S_m$ (m=1, 2, …6). We use the same 6 polarization states demonstrated in Fig. 2b in the simulation, and calculate the 3D super-resolution spatial distribution of the polar and azimuthal angles from the reconstructed 3D super-resolution image of $S_m$ (m=1, 2, …6).

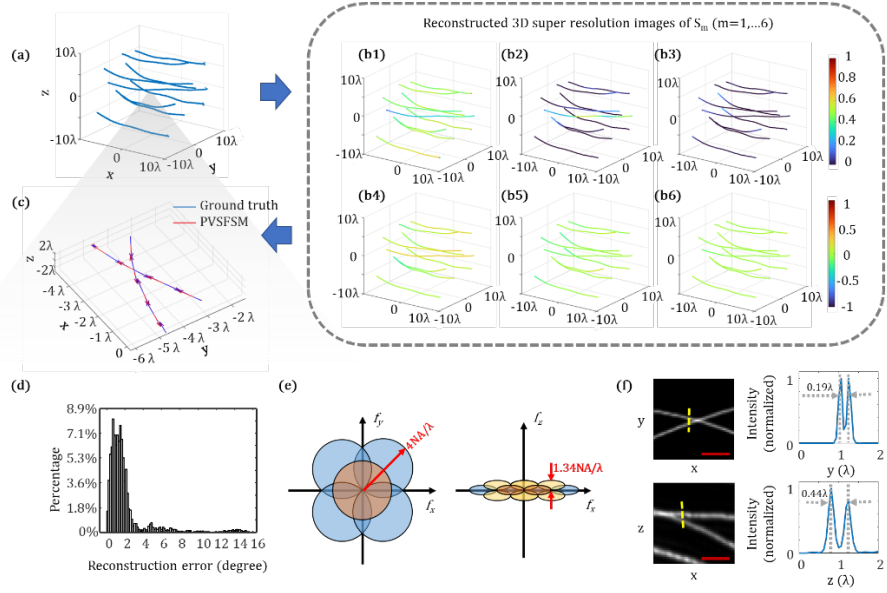

**Fig. 5.** 3D-orientation-3D-PVSFSM imaging. **(a)** Ground truth distribution of microfilament. **(b1-b6)** The reconstructed 3D super-resolution images of six polarization components of $S_m$ (m=1, 2, …6). **(c)** 3D vector plot of the reconstructed and real 3D orientation of a magnified region in (a). **(d)** The statistical reconstruction error of 3D orientation angle. Statistical analysis was performed on 2247 points. **(e)** The detected spatial spectrum ranges in the x-y and x-z perspective view. **(f)** Lateral and axial line scans of the image across two adjacent microfilaments. λ in the figure denotes $\lambda^{det.}$. NA = 1.4. $\lambda^{det.}$ = 640 nm.

In the simulation of 3D-orientation-3D-PVSFSM, a microfilament structure decorated with fluorescent molecules oriented parallel to it is utilized as an object, as shown in Fig. 5a. In M(***r***), the α is set to 67 degrees to match the objective NA (NA = 1.4) to enable a two times resolution improvement in both the lateral and axial dimension. Fig. 5b1-b6 shows the reconstructed 3D super-resolution image of $S_m$ (m=1, 2, …6). With a Gaussian-Newton optimization method, the polar and azimuthal angular information about the 3D orientation have been calculated from $S_m$. Fig. 5c shows the 3D vector plot of the 3D orientation of a magnified view, in which the reconstruction result coincides well with the ground truth. A statistical analysis performed on 2247 dipoles demonstrates an angular deviation of 2.01±2.49 degrees, as shown in Fig. 5d. Fig. 5e shows the expanded detection range in the reciprocal domain. Lateral and axial line scans of the image across two adjacent microfilaments demonstrate the lateral and axial resolutions of $0.19\lambda^{det.}$ and $0.44\lambda^{det}$ (Fig. 5f).

*4.2 Speed improvement*

The imaging speed of PVSFSM is lower than the conventional spatial-frequency-shift microscopy as the structured illumination pattern is generated by a scanning process rather than a wide-field projection. For example, the scanning process would take about 4 seconds to cover a 10 μm×10 μm field of view (considering 100 nm step size to meet the Nyquist condition and a typical response time of 400 μs for a Galvo scanning mirror). Then for the result demonstrated in Fig.2, it costs 7 minutes to capture all the raw images [corresponding to (6 FSVs)×(3 phases)×(6 polarization states)]. To speed up the scanning process, we can add a spatial light modulator in the illumination light path to generate multiple foci for parallel illumination[29]. The adaptive optics can also be used to skip the position where no signal (i.e. blank region of the sample) comes out[30].

It is also possible to reconstruct the 3D orientation using the polarization-modulated wide-field spatial-frequency-shift microscopy, to improve the imaging speed by escaping from the scanning process. According to our deduction, the captured image can be expressed by Eq. 4. While, it is difficult to decompose the spatial frequency spectrum of the captured image into several sub-components in the reciprocal domain regarding the functions of $\theta$ and $\varphi$, as in 3D-orientation PVSFSM demonstrated above, which poses more complexity on the reconstruction. Reconstructing the 3D orientation distribution directly in the spatial domain is an alternate way[31, 32]. This is beyond the scope of this manuscript, and we may study the reconstruction method in 3D orientation super-resolution imaging via wide-field structured illumination in the future.

$$I = \left\{ \begin{matrix} A_p \left[ \cos(2\pi f_0 x + \varphi_p) \cos(\varphi - \varphi_0) \sin\theta + \tan\theta_0 \sin(2\pi f_0 x + \varphi_p) \cos\theta \right] \\ + A_s \cos(2\pi f_0 x + \varphi_s) \sin(\varphi - \varphi_0) \sin\theta \end{matrix} \right\}^2 S \otimes PSF^{det.} \qquad \textbf{Eq. 4}$$

where $A_{p/s}$, $\varphi_{p/s}$, and $2\pi f_0$ denote the amplitudes (subscript p for p polarized component and subscript s for s polarized component of the obliquely incident plane wave for illumination), the pattern phases, and the lateral wave number of the interfered standing wave. $\varphi_0$ and $\theta_0$ denote the azimuthal and polar angle of the obliquely incident plane wave. S describes the emission rate of the sample with polarization property. The emission rate of the sample without polarization property is ignored for simplicity.

## 5. Conclusion

In conclusion, we propose a novel PVSFSM for high-throughput and universal super-resolution mapping of the 3D orientation. The structured illumination pattern for introducing the spatial-frequency-shift effect is generated by rapidly scanning an intensity-modulated tight focus in the sample domain. The spatial-angular distribution mixed in the real space can be well separated into several sub-components in the reciprocal space regarding 6 different functions of the orientation angular parameters, which can be easily solved with a well-conditioned coefficient matrix provided by switching 6 different polarization states in the back aperture of the objective. Then the 3D orientation angular information can be reconstructed with super resolution by stitching the solved sub-components in the reciprocal domain and following a Gaussian-Newton optimization calculation. 3D orientation mapping with lateral resolutions of $0.19\lambda^{det.}$ and $0.25\lambda^{det.}$ has been demonstrated for sparsely distributed dipoles in noise-free and heavy-noisy (13dB SNR) cases respectively (NA=1.4). The ability to extract 3D spatial distribution and 3D orientation simultaneously has also been illustrated, with the lateral and axial resolution both two times improved from the Abbe diffraction limit. The proposed method overcomes the retractions on the current 3D orientation super-resolution microscopy such as the requirement of special labeling or low throughput caused by capturing thousands of raw frames, which could broadly benefit life science research, bio-medicine, and material science, among others.

**Acknowledgments.** National Natural Science Foundation of China (No. T2293751, 62020106002, 92250304, 61905097, 62005250); Department of Science and Technology of Zhejiang Province "Leading Goose" program (No. 2022C01077); National Key Basic Research Program of China (grant no. 2021YFC2401403 and 2022ZD0119400).

**Disclosures.** The authors declare no conflicts of interest.

**Data availability.** The data that support the findings of this study are available from the corresponding author upon reasonable request.